\documentclass[fleqn,twoside]{article}
\usepackage{espcrc2}
\usepackage{graphicx}
\usepackage[figuresright]{rotating}

\newcommand{\AmS}{{\protect\the\textfont2
  A\kern-.1667em\lower.5ex\hbox{M}\kern-.125emS}}
\newcommand{\pbp}{\langle\bar{\psi}\psi\rangle}

\title{Diquark condensation in dense SU(2) matter}

\author{S. Hands\address[Swan]{Department of Physics, University of Wales Swansea,
	 Singleton Park, Swansea SA2 8PP, UK},
	I. Montvay\address[DESY]{Theory Division, DESY, Notkestra{\ss}e 85, D-22603 Hamburg, 
	Germany},
	L. Scorzato\addressmark[Swan]\addressmark[DESY],
	J. Skullerud\addressmark[DESY]\address[Ams]{ITF, Universiteit van Amsterdam, 
	Valckenierstraat 65, 1018 XE Amsterdam, The Netherlands}}
\begin{document}

\begin{abstract}
We report on a lattice study of two-color QCD with adjoint staggered fermions at high density.
We find that the model has no early onset and we report on results for diquark condensation, 
from simulations with and without a diquark source term. 
\vspace{1pc}
\end{abstract}

\maketitle

\section{INTRODUCTION}
Numerous model calculations suggest that the ground state of QCD at high
baryonic density is characterized by diquark condensation which spontaneously
breaks gauge and/or baryon number symmetry (see \cite{Rajagopal:2000wf,Alford:2001dt}
for recent reviews). In \cite{Hands:2000ei,Hands:2000yh} we presented evidence
that two-color QCD with one flavor of adjoint staggered fermions 
has interesting QCD-like features, in particular the absence of baryonic Goldstone
modes in the spectrum. Here we extend the study of this model to include
possible diquark condensates. We refer to \cite{Hands:2000ei} for a detailed
description of the model and its properties, while we summarize here only the main
features: $N$ flavors of staggered fermion, in the adjoint representation of the
SU(2) gauge group, have a U(2$N$) chiral-flavor symmetry (at zero mass $m$ 
and zero chemical potential $\mu$). This is explicitly 
(spontaneously) broken by the mass (the chiral condensate) to Sp(2$N$). 
SSB gives rise to $2N^2-1$ Goldstone modes, of which $N(N-1)$ are baryonically charged.
When $N>1$ and $\mu> m_\pi/2$ the vacuum starts rotating from the chiral condensate
direction to that of a diquark condensate \cite{Kogut:2000ek}, which in the case
of $N=2$ is (modulo a baryonic rotation):
\[
qq_3=\frac{i}{2}\left[\chi^{p\,tr}(x)\varepsilon^{pq}\chi^q(x)+
\bar\chi^p(x)\varepsilon^{pq}\bar\chi^{q\,tr}(x)\right],
\]
where $\varepsilon^{pq}, \; p,q\!=\!0,1$ is the completely anti-symmetric 
tensor in flavor space. If $N=1$ no baryonic Goldstone mode is expected, 
and no local, gauge invariant, Lorentz scalar diquark condensate is possible.
Therefore no early onset ($\mu\sim m_\pi$) transition is expected.
One interesting possibility is a gauge non singlet --- hence color 
superconducting --- diquark condensate:
\[
qq_{sc}^i=
\frac{1}{2}\left[\chi^{tr}(x)t^i\chi(x)+\bar\chi(x)t^i\bar\chi^{tr}(x)\right],
\]
where $t^i, \; i=1,3$ are color generators. Note however that the fermionic
determinant of the $N=1$ model (although real) does not have a definite sign. 
The indication in \cite{Hands:2000ei,Hands:2000yh} was that the effect of the
sign is indeed important. If we ignore the sign, the model is well described by
Chiral Perturbation Theory ($\chi$PT) \cite{Kogut:2000ek}, 
including the presence of baryonic Goldstone modes.
The latter are suppressed only by the inclusion of the sign. Here we want to complement
the previous analysis with the study of diquark condensation \cite{Hands:2001}.
\section{ALGORITHMS}
We studied the $N=1$ model by means of both HMC \cite{Duane:1987de} and 
TSMB \cite{Montvay:1996ea} algorithms. Because of lack of ergodicity, with HMC
we could explore only the positive determinant phase. TSMB is defined by
a set of parameters (degrees of polynomia $n_i$, the number of heatbath and 
overrelaxation iterations for the bosonic fields $I_h$ and $I_o$, the number of metropolis
iterations for the gauge fields $I_M$ and the number of noisy corrections $I_c$).
One has to tune such parameters in order to minimize the autocorrelation.
The latter must be expressed in term of matrix multiplications. 
The number of matrix multiplications per
sweep is roughly given by:
\[
N_{mult} \simeq \frac{7}{2}n_1(I_h+I_o+I_M) + (n_2+n_3)I_c.
\]
\begin{table}
\caption{Runs at $\beta=2.0,\; m=0.1,\,$ on a $4^3\times 8$ lattice.}
\label{table:autocorr}
\begin{tabular}{rrrrrrrr}
\hline
 $\mu$ & Alg. & $n_1$ & $\lambda/\epsilon$ &  $N_{sw}$  &   $N_{int}^{plaq}$  \\
\hline
 $0.0$ & {\tiny TSMB} & 24 & $2 \cdot 10^3$ & $1 \cdot 10^5$ & $2.1 \cdot 10^5$ \\
 $0.0$ & {\tiny TSMB} & 16 & $2 \cdot 10^3$ & $2 \cdot 10^5$ & $1.7 \cdot 10^5$ \\
\hline
 $0.0$ & {\tiny HMC} & --- & $2 \cdot 10^3$ & $3 \cdot 10^4$ & $1.9 \cdot 10^5$ \\
\hline
 $0.30$ & {\tiny TSMB} & 48 & $8 \cdot 10^4$ & $1 \cdot 10^5$ & $9.2\cdot 10^6$ \\
 $0.37$ & {\tiny TSMB} & 80 & $4 \cdot 10^5$ & $1 \cdot 10^5$ & $2.5\cdot 10^7$ \\
\hline
\end{tabular}\\[2pt]
$N_{int}^{plaq}$ is the autocorrelation length (expressed in number of 
matrix multiplications) for the plaquette, evaluated from a run of $N_{sw}$ sweeps.
$n_2=90$ at $\mu=0$, otherwise $n_2\sim 10 n_1$.
\end{table}
\begin{table}
\caption{Runs at $\beta=2.0,\; m=0.1,\,$ on a $4^3\times 8$ lattice.}
\label{table:runparam}
\begin{tabular}{rrrrrrr}
\hline
$\mu$ & $n_1$ & $n_2$  & $N_g$ & $p_-$ & $\langle r\rangle$ \\ \hline
0.00 &  16 &  120 & 380 & 0.00 & 1.00 \\
0.30 &  48 &  500 & 216 & 0.00 & 0.9982(3) \\ 
0.36 &  64 &  700 & 140 & 0.14 & 0.476(14) \\
0.37 &  80 &  800 & 275 & 0.22 & 0.45(2) \\
0.38 & 100 & 1000 & 440 & 0.33 & 0.30(4) \\
0.40 & 100 & 1000 & 265 & 0.44 & 0.085(24)\\ \hline
\end{tabular}\\[2pt]
$N_g$ is the effective number of {\em independent} configurations.
$p_-$ is the fraction with a negative
determinant and $\langle r\rangle$ the average reweighting factor including
the sign. 
\end{table}
In general when the chemical potential increases the lowest eigenvalues ($\epsilon$)
approach zero and the simulations become difficult.
In Table~\ref{table:autocorr} we show how the condition number ($\lambda/\epsilon$, 
$\lambda$ being the largest eigenvalue)
as well as the plaquette autocorrelation change by increasing $\mu$. 
The autocorrelation of fermionic observables is typically about 10 times
shorter.
From  Table~\ref{table:runparam} we see that until the point $\mu=0.38$ the average
sign is still not too small, while points above $\mu=0.4$ are already exceedingly difficult.
\section{DIQUARK CONDENSATES}
Our model
has very interesting features, which distinguish it from other SU(2) models. 
It also has a sign problem. It is natural and interesting to study the effect of 
the inclusion of the sign.
We distinguish two models: the one where only positive (determinant) 
configurations are sampled 
and the one where both positive and negative 
configurations are included with their sign.
The positive sector turns out to be in very good agreement with $\chi$PT 
predictions; i.e., the positive model behaves effectively like an $N=2$ model.

\begin{figure}[htb]
\includegraphics[width=0.45\textwidth,height=0.45\textwidth]{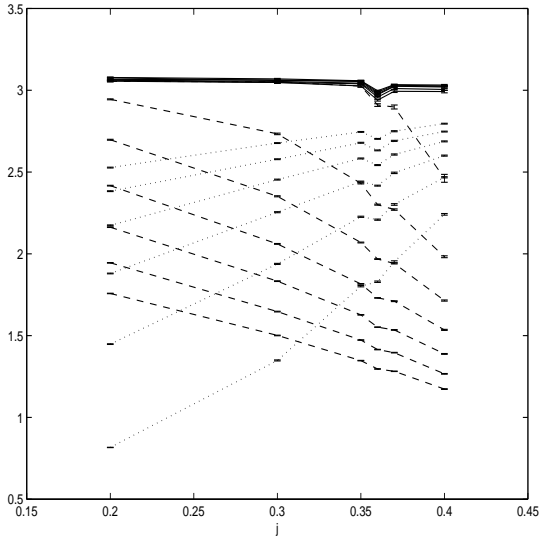}
\caption{Dotted (increasing) lines are $qq_3$, dashed (decreasing) lines 
are $\pbp$ (at different values of $\mu$ between 0 and 0.5). Solid lines are the sum in 
(\ref{eq:sum}). The error bars do not take into account the autocorrelation.}
\label{fig:sumjmu}
\end{figure}

In order to study possible diquark condensations we introduced a source $j$ for
$qq_3$. This is done in a partially quenched way: by introducing the 
source only in the measurements.
$\chi$PT predicts also the dependence on $j$ of the chiral
and the diquark condensate \cite{Kogut:2000ek}. In particular the sum
\begin{equation}
\label{eq:sum}
\pbp_0^2=\pbp^2+\langle qq_3\rangle^2
\end{equation}
must, to lowest order in $\chi$PT, be a constant independent of 
both on $\mu$ and $j$ (for NLO see \cite{Splittorff:2001fy}). In fact, as one
can see on fig.~\ref{fig:sumjmu}, this relation is very well satisfied
in the positive sector.
The validity of the relation (\ref{eq:sum}) also allows an extrapolation 
of $\langle qq_3\rangle$ to zero $j$, which would otherwise be quite
difficult. 
\begin{figure}[htb]
\includegraphics[width=0.45\textwidth,height=0.45\textwidth,angle=-90]{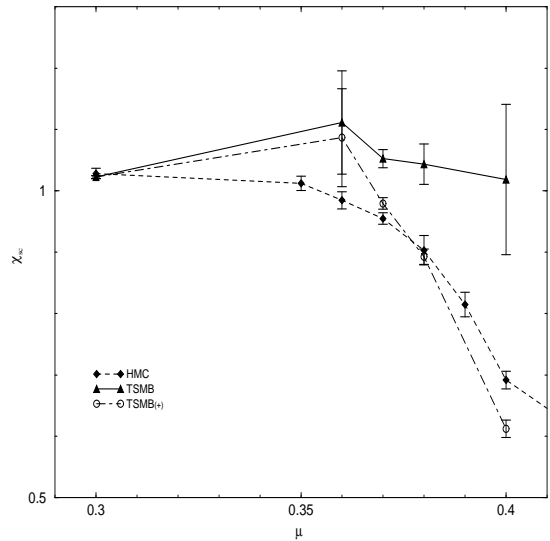}
\caption{Superconducting susceptibility as a function of $\mu$. 
The effect of including the sign is compared with the positive sector.}
\label{fig:subpbp}
\end{figure} 

Since the condensate $qq_{sc}$ is not gauge invariant we studied its 
susceptibility $\chi_{sc}$
\[
\chi_{sc}  = \frac{1}{3}\langle\bar\chi(x) t^i\bar\chi^{tr}(x)
 \chi^{tr}(x)t^i\chi(x)\rangle \, . 
\]
This has a finite value even without a source term. In the low
density regime $\chi_{sc}$ shows little variation with $\mu$. Once we
reach the $\chi$PT transition point, $\chi_{sc}$ drops rapidly 
in the positive sector. We interprete this as an effect of Pauli blocking or phase space
suppression: as the ground state is filled up by fermions there is
less phase space left to accommodate 
the fermion loops that contribute to $\chi_{sc}$.
When the sign is properly taken into account this effect disappears and 
$\chi_{sc}$ remains stable through the transition (fig.~\ref{fig:subpbp}).

\section{CONCLUSIONS}
The interesting lesson one can learn from this model is that the early onset
transition, typical of all other SU(2) models, is indeed delayed as
expected from the analysis of possible diquark condensates. Numerically
this is obtained by a delicate balance between configurations with
positive and negative determinant. This
makes simulations prohibitively expensive in the region where one would expect 
a true onset transition at high density. 


This work is supported
by the TMR network ``Finite temperature phase transitions in particle physics''
EU contract ERBFMRX-CT97-0122. 
Numerical work was performed using a Cray T3E
at NIC, J\"ulich and an SGI Origin2000 in Swansea.


\end{document}